\documentclass[aps,pra,twocolumn, showpacs]{revtex4-1}
\usepackage{amsmath, amsthm, amssymb,amsbsy,graphicx,times,subfigure,marvosym,color,bm,hyperref}
\usepackage[latin1]{inputenc}
\usepackage{sidecap}

\usepackage{ulem}

\begin{document}

\title{Synchronization transition in dipole-coupled two-level systems with positional disorder}
\author{M. P. Kwasigroch}
\author{N. R. Cooper}
\affiliation{T.C.M. Group, Cavendish Laboratory, University of Cambridge, J. J. Thomson Avenue, Cambridge CB3 0HE, U.K.}


\begin{abstract}

We study the decoherence dynamics of dipole-coupled two-level quantum systems in Ramsey-type experiments. We focus on large networks of two-level systems, confined to two spatial dimensions and with positional disorder giving rise to disordered dipolar couplings. This setting is relevant for modeling the decoherence dynamics of the rotational excitations of polar molecules confined to deep optical lattices, where disorder arises from the random filling of lattice sites with occupation probability  $p$. We show that the decoherence dynamics exhibits a phase transition at a critical filling $p_c\simeq 0.15$.  For $p<p_c$ the dynamics is disorder-dominated
and the Ramsey interference signal decays on a timescale  $T_2 \propto p^{-3/2}$. For $p>p_c$ the dipolar interactions dominate the disorder, and the system behaves as a collective spin-ordered phase, representing synchronization of the two-level systems and persistent Ramsey oscillations with divergent $T_2$ for large systems. For a finite number of two-level systems, $N$, the  spin-ordered phase at $p> p_c$  undergoes a crossover to a collective spin-squeezed state on a timescale $\tau_{\rm sq} \propto \sqrt{N}$.
We develop a self-consistent mean-field theory that is capable of capturing the synchronization transition at $p_c$, and  provide an intuitive theoretical picture that describes the phase transition in the long-time dynamics. We also show that the decoherence dynamics appear to be ergodic in the vicinity of $p_c$,  the long-time behaviour being well described by the predictions of equilibrium thermodynamics. The results are supported by the results of exact diagonalization studies of  small  systems.

\end{abstract}



\date{22 August, 2017}

\maketitle


\section{Introduction}

The dynamical evolution of coupled quantum spin systems is a very rich topic for exploration with a long and fruitful history, starting with early works on the dynamics of nuclear spins.  The considerations of the motion of a single spin-flip in a disordered network of coupled nuclear spins led Anderson to propose of the notion of localization of single particles via quantum interference~\cite{andersonlocalization}, and to point out important questions concerning the qualitative form of  dynamics in the many-particle case.

The dynamics of  coupled quantum spin systems is undergoing a resurgence of  interest, due to recent advances in both theory and experiment. The theoretical advances include  the development of the concept of many-body localization (MBL)\cite{Huse2013}, involving the non-ergodic behaviour of isolated many-body quantum systems, for which models of highly excited quantum spin systems provide an important class of examples. The experimental advances involve the development of numerous novel forms of coupled two-level quantum systems, which constitute generalized quantum ``spins",  in which the rate of  decay of excitations ($1/T_1$ relaxation rate) is small compared to the timescales set by the coupling between the spins. These systems include not just nuclear spins, but also trapped ions, NV centres, Rydberg excitations, and polar molecules\cite{ions,nv,rydberg,polarmols}. They provide a range of novel settings  (different dimensionalities,  range and nature of coupling, and forms of  disorder) in which collective quantum spin dynamics can be studied. 

In recent work~\cite{kwasigrochcooper} we have shown that,  in certain settings,
simple Ramsey-type experiments on quantum spin systems with transverse (XY) spin-spin couplings can show an interesting phase transition in
  the long-time dynamics: between a phase that is disorder-dominated and decoheres at long times (finite $T_2$ decoherence time); and a phase where dipole interactions dominate the disorder and give rise to a collective spin-ordered state that retains coherence for arbitrarily long times ($T_2\to \infty$). This long-lived coherence indicates that the phase describes a synchronized, phase-locked, oscillation of all of the two-level systems.
The synchronization transition proposed in  Ref.~\onlinecite{kwasigrochcooper} arises in a closed quantum system  (under unitary time evolution and conserved energy) so is distinct from synchronization phenomena in driven open quantum systems\cite{zhu} or in classical models of driven dissipative XY systems\cite{wiesenfeld,chowdhurycross}.

This collective phase-locked oscillation of all the two-level systems is a consequence of certain special features that arise for dipolar coupling in a two-dimensional (2D) plane. These features can be understood using intuition based on the thermodynamic phases of hard-core bosons, representing the $s=1/2$ quantum spins in a particle-like picture\cite{kwasigrochcooper}. Within this picture, the XY spin-spin coupling leads to inter-site hopping of the bosons, which has interesting features for dipolar interactions in 2D.  On the one hand,  a hopping amplitude of   $1/r^\alpha$ with $\alpha=3$ is sufficiently {\it short-ranged} in $D=2$ dimensions   ($\alpha > D$) to permit a well-defined thermodynamic limit with a finite energy density. (This contrasts with dipolar interactions in $D=3$ for which surface effects, such as sample shape, can control bulk behaviour.) It is also sufficiently short-ranged to allow the existence of localized single-particle states in the regime of strong disorder~\cite{andersonlocalization}. On the other hand, for a disorder-free system, this form of hopping  is sufficiently {\it long-ranged}  in 2D  ($\alpha <  2D$) to allow the existence of a Bose-Einstein condensate (BEC) of these particles -- {\it i.e.} with long-ranged phase coherence --  even at non-zero temperature. Specifically, for
$\alpha=3$ in $D=2$ dimensions, the long-wavelength dispersion of the bosons is relativistic: this leads to a density of states that vanishes at the band edge,  allowing a stable BEC of bosons even at non-zero temperatures in 2D~\cite{buechler}. (This contrasts with the usual case of parabolic dispersion, for which there is no such BEC in $D=2$.) Thus, for a sufficiently clean system, one anticipates that the bosons can retain long-range phase coherence even for excited states (sufficiently close in energy to the ideal phase-ordered ground state) reflecting the existence of this stable BEC phase of hard-core bosons even for nonzero temperatures. 
Indeed, in Ref.~\onlinecite{kwasigrochcooper} we showed theoretical evidence for the existence of this phase in 
a quantum spin system excited by a uniform pulse (as in a standard Ramsey interference experiment), and for its destruction by strong disorder arising from inhomogeneous broadening of the individual two-level transitions. 

In the present paper, we investigate the properties of this collective spin-ordered phase in the case where there is  only  {\it positional} disorder, and no local frequency offsets. This is the situation most relevant to the quantum spin systems formed by  polar molecules in deep optical lattices.  There, the two levels are two rotational levels of the polar molecule (e.g. ro-vibrational ground state and one rotationally excited level~\cite{Micheli2006,Lukin2006, Buchler2007, Wall2009, Watanabe2009, 
Yu2009, Wall2010,Trefzger2010,Krems2010,
  Daley2010,Kestner2011,GorshkovPRL,GorshkovPRA}). The positional disorder arises from the incomplete, and random, site occupancy of the lattice sites. To describe this situation requires one to go beyond the simple mean-field models used in Ref.~\onlinecite{kwasigrochcooper}. We do so by developing a self-consistent mean-field theory involving an RPA-like decoupling of correlation functions. We present the results of this model for the long-time dynamics, and demonstrate the existence of a phase transition at site occupation probability of $p_c \simeq 0.15$.
We further show that this is close to the transition expected under the assumption that the dynamics are ergodic, so can be interpreted in terms of the equilibrium phase transition. We present results of exact diagonalization studies of small systems which confirm this general picture.

Throughout this work, we present our results in the language of quantum spin systems, rather than of hard-core bosons used in Ref.~\onlinecite{kwasigrochcooper}.  In terms of spins, the BEC of hard-core bosons of Ref.~\onlinecite{kwasigrochcooper} can be viewed as a collectively ordered spin state, with long-range spin-spin correlations. This long-range ordering of the spins translates to a synchronized, phase-locked, oscillation of widely-separated two-level systems.
We also draw closer connections to the physics of Ramsey-type experiments on two-level quantum systems. In particular, we show that in system of finite number of spins $N$, the dynamics of the collective spin-ordered phase shows a crossover to a strongly spin-squeezed state on a timescale $\tau_{\rm sq} \propto \sqrt{N}$. 
\section{Model} 

  We study a set of coupled two-level quantum systems, which we shall view as $s=1/2$ spins, with Hamiltonian
\begin{equation}
H=\frac{J_0 a^3}{2}\sum_{i\neq j}\frac{1}{|\mathbf{R}_i-\mathbf{R}_j|^3}(s^-_is^+_j+s^-_js^+_i),
\label{eq:ham}
\end{equation}
where $s^{\pm}_i$ are the raising/lowering operators for the $i$-th spin, which is located at  position ${\bm R}_i$. (We study the spin system in a frame rotating at the bare transition frequency $\omega_0$, which is assumed to satisfy $\omega_0 \gg J_0/\hbar$ such that the rotating wave approximation is accurate.) The positions ${\bm R}_i$ are assumed to lie in a 2D plane at the sites of a square lattice of lattice constant $a$. Crucial to our study, however, is the assumption that each site hosts a spin only with (independent) probability $p$. Thus, the mean 2D density of spins is $p$ (in units of the density of sites), and for $p\neq 1$ the system has (quenched) positional disorder. 

The model is relevant in a number of dfferent physical settings in which two-level quantum systems are coupled by dipolar interactions, including polar molecules,  NV centres, and coupled Rydberg excitations.  Note, however, that the specific physical setting could lead to variants of this model --  in dimensions other than $D=2$, or including local frequency offsets (random fields that couple to $s^z_i$),  or additional $s^z_is^z_j$ interactions\cite{Baranov2012} -- which could lead to differences in the qualitative physics that we describe.

Our primary motivation has been to understand the coupled rotational excitations of polar molecules confined to the sites of an optical lattice~\cite{ColoradoExp}, for which this model emerges naturally~\cite{kwasigrochcooper}. In that case, we can take  $|s^z=-1/2\rangle$ to represent the ro-vibrational ground state, $|\ell=0,m=0\rangle$, of the polar molecule at site $i$ and  $|s^z=+1/2\rangle$  the $|\ell=1,m=0\rangle$ rotationally excited state ($\ell$ is the molecular angular momentum and $m$ its projection perpendicular to the 2D plane). Then the Hamiltonian (\ref{eq:ham}) describes the  resonant ``flip-flop" transfer of a rotationally excited state between pairs of molecules driven by the dipolar coupling~\cite{Lukin2006}. The interaction energy $J_0= d^2/4\pi\epsilon_0 a^3$ is the coupling strength for nearest neighbour sites in
terms of the dipole matrix element $d$ between $|\ell=0,m=0\rangle$, and  $|\ell=1,m=0\rangle$ (e.g. $J_0/h = 52$~Hz for
KRb in a lattice with $a = 532$ nm~\cite{ColoradoExp}).  Although lattices of polar molecules are typically prepared as 3D lattices, the dynamics of the rotational excitations could be confined and studied in 2D by use of an electric or magnetic field gradients to isolate the flip-flop resonance condition to a 2D plane. 
Technical advances have dramatically increased the available filling of lattice sites with ground state polar molecules above $p\simeq 0.3$~\cite{lowentropymolecules,innsbruck}. As we will describe, our results show that this upper limit is sufficient to show an interesting phase transition to an interaction-stabilized synchronized phase that is resistant to decoherence, and which shows spontaneous spin-squeezing.

\subsection{Method of Probing and Physical Observables}

We consider probing the system by means of Ramsey interference. 
Starting from the de-excited state $\prod_{i=1}^N |s^z=-1/2\rangle_i$,
we consider applying a resonant pulse with uniform amplitude and phase, to prepare the initial state
\begin{equation}
|\Psi\rangle=\prod_{i=1}^N\left(\cos\frac{\theta}{2} |s^z=-1/2\rangle_i + \sin\frac{\theta}{2} e^{i\phi} |s^z=+1/2\rangle_i\right)
\label{eq:prepare}
\end{equation}
Here the site label $i$ runs over all sites on which there are spins, which is typically $N = p N_{\rm site}$ for a 2D lattice of $N_{\rm site}$ sites.
The initially prepared state (\ref{eq:prepare}) has  a uniform density $\rho = p \sin^2({\theta}/{2})$ of excited states, and with full  phase coherence between lattice sites.
We then study the subsequent temporal dynamics under the action of (\ref{eq:ham}).  Our goal is to
determine the long-time behaviour.  Since the Hamiltonian (\ref{eq:ham}) conserves total $s^z$, the mean density of excited states,    $\rho$, must be conserved. The central question is: does the
system retain long-range spin-spin correlations? 

We shall show that for large systems, $N\gg 1 $, there is a phase transition between a disorder-dominated phase at $p<p_c$ in which  individual two level systems decohere from each other at long times, and a spin-ordered phase at $p>p_c$  in which the  dipole-dipole interactions cause the spins to remain collectively locked together to arbitrarily long time.  
This collective phase has striking signatures in the  Ramsey interference signal, obtained by applying the conjugate pulse to rotate the spins back by angle $-\theta$. For a  large system, $N\to \infty$,  the signature of this collective phase is a diverging $T_2$ decoherence time (i.e. persistent Ramsey fringes). For a finite system, there is a crossover, on a timescale $\tau_{\rm sq}$ that grows as $\sqrt{N}$, to a collective state of the spins which exhibits spin-squeezing~\cite{kitagawaueda,cappellarolukin}.

We shall study the dynamics of the initial state with $\theta= \pi/2, \phi=0$, with all spins in  $s_x=1/2$, which is representative of the general case. It is convenient to rotate coordinates to define the new spin operators  $S^x = -s^z$, $S^y = s^y$, $S^z = s^x$, such that the initial state (\ref{eq:prepare}) is the product of eigenstates of $S^z=+1/2$. In this frame the Hamiltonian~(\ref{eq:ham}) becomes
\begin{eqnarray}
H =\sum_{i\neq j} J_{ij} \left[S^z_iS^z_j-\frac{1}{4}(S^+_i-S^-_i)(S^+_j-S^-_j)\right] \label{Ham}
\end{eqnarray}
where $J_{ij} \equiv {J_0 a^3}/{|\mathbf{R}_i-\mathbf{R}_j|^3}$.
The prepared state is an eigenstate of the first term on the right hand side of Eq.~\ref{Ham}. We shall study the dynamics induced by the remaining terms. 
Our primary focus will be on the time evolution of the
mean polarization 
\begin{equation}
\langle S^z\rangle \equiv (1/N)  \sum_i \langle S^z_i\rangle\,,
\end{equation} with $N$ the total number of spins. This quantity $\langle S^z\rangle$ determines the amplitude of the Ramsey interference signal (i.e. the oscillations at the transition frequency $\omega_0$). However, for reasons we expand on below, it will be helpful also to study the quantity
\begin{equation}
\Delta  \equiv  \frac{1}{N^2}\sum_{i \neq j} \langle S^z_i S^z_j +S^y_iS^y_j \rangle \label{eq:rho0},
\end{equation} 
This describes the mean-square of the collective spins $S^z \equiv (1/N)  \sum_i S^z_i$
and $S_y \equiv (1/N)  \sum_iS^y_i$ (up to an offset of $1/2N$). It therefore acts as an order parameter for the long-range coherence of the spin. This order parameter is helpful in exposing the existence of a long-range spin-ordered state, $\Delta \neq 0$, even in regimes for which the mean polarization vanishes, $\langle S^z\rangle =0$, arising from the formation of a spin-squeezed state of the collective spin~\cite{cappellarolukin}. The order parameter $\Delta$ could be  obtained from Ramsey-type experiments by measuring the {\it variances} of the distributions of $S^z$ and $S^y$ obtained in repeated measurements. 
[Note that,  when the individual two-level systems are viewed as hardcore bosons, $\Delta$ describes the {\it condensate fraction} of these bosons. Thus,
the BEC phase of hardcore bosons described in Ref.~\cite{kwasigrochcooper} is equivalent to the collective spin-ordered phase, $\Delta \neq 0$.]

\section{Self-Consistent Mean Field Theory}

A complete description of the dynamics requires a solution of the full
time-evolution of this disordered many-particle quantum spin
system. Since this is too complex for an exact solution\cite{schachenmayer}, we develop an
effective mean-field description that is qualitatively and
quantitatively accurate.  In the simplest mean-field theories, in
which the Hamiltonian is reduced to a linear coupling of spins to the
mean polarization $\langle S^z\rangle$, the dynamics becomes trivial
even for a disordered lattice, $p\neq 1$: the mean field polarization
$\langle S^z\rangle$ remains time independent.  (This is a consequence
of the fact that disorder enters only in the magnitudes of the
couplings $J_{ij}$, so a mean-field state with uniform
$\langle S^z_i\rangle$ remains an extremum of the energy.) The system
only evolves in time if quantum fluctuations about this simplest
mean-field theory are included.
 
\subsection{RPA decoupling}

To develop a theory that goes beyond the simplest mean-field theory,
we study the dynamics in a form of random phase approximation (RPA)
known as Tyablikov decoupling~\cite{Tyablikov}. This takes into
account the leading quantum fluctuations above the mean-field state in
a self-consistent manner. The approach involves re-expressing all
3-point correlation functions that enter the equations of motion in
terms of 1- and 2-point ones, e.g.
$\langle S^z_i S^+_j S^-_k \rangle \simeq \langle S^z_i\rangle \langle
S^+_j S^-_k\rangle$.  We obtain a closed system of coupled equations
for $\langle S^z_i \rangle$ and the equal-time 2-point correlation
functions $\langle S^+_iS^+_j\rangle$, $\langle S^-_iS^-_j\rangle$ and
$\langle S^+_iS^-_j\rangle$:
\begin{eqnarray}
\frac{d}{dt}\langle S^z_a\rangle &=&-\sum_{i}2J_{ia}\langle S^y_iS^x_a\rangle,\nonumber\\
\frac{d}{dt}\langle S^y_aS^x_b\rangle&=&\sum\Big[2\langle S^z_i\rangle\left(-J_{ia}\langle S^x_aS^x_b\rangle +J_{ib}\langle S^y_a S^y_b\rangle\right)\nonumber\\
&&-2J_{ib}\langle S^z_b\rangle \langle S^y_a S^y_i\rangle\Big]
=\frac{d}{dt}\langle S^x_b S^y_a\rangle,\nonumber\\
\frac{d}{dt}\langle S^y_aS^y_b\rangle &=& \sum2\langle S^z_i\rangle\left(J_{ia}\langle S^x_aS^y_b\rangle + J_{ib}\langle S^y_aS^x_b\rangle\right)\nonumber\\
\frac{d}{dt}\langle S^x_aS^x_b\rangle &=&\sum\Big[-2\langle S^z_i\rangle\left(J_{ia}\langle S^y_aS^x_b\rangle + J_{ib}\langle S^x_aS^y_b\rangle\right)\nonumber\\
&&+2J_{ia}\langle S^z_a\rangle \langle S^y_i S^x_b\rangle +2J_{ib}\langle S^z_b\rangle\langle S^x_aS^y_i\rangle\Big].\label{system}
\label{eq:theory}\end{eqnarray}
The above system of equations can be generated from a Hamiltonian in which longitudinal (massive) fluctuations are neglected:
\begin{eqnarray}
H &=&\sum_{i\neq j} J_{ij} \left[S^z_iS^z_j + S^y_iS^y_j\right] 
\nonumber\\
&=&
\sum_{i\neq j} J_{ij} \Big[\langle S^z_i\rangle \langle S^z_j \rangle + 2 \left( S^z_i - \langle S^z_i\rangle \right) \langle S^z_j \rangle+ S^y_iS^y_j
\nonumber\\
&& + \mathcal{O} \left((S^z_i - \langle S^z_i\rangle) (S^z_j - \langle S^z_j\rangle)\right)\Big] .
\end{eqnarray}

\subsection{Analytic Theory}

We now describe the main structure of the theory and the resulting dynamical transition captured by the above equations. To simplify the presentation, and to allow some analytical insights, in this section we shall make a coarse-graining approximation $\langle S^z_i \rangle \simeq \langle S^z \rangle$. (Comparisons with numerical results, described below, show that this coarse-graining procedure reproduces the full results accurately.) A reader most interested in the final results may skip over this subsection on theoretical methods to the numerical results in subsection~\ref{numerical}.

In the coarse-graining approximation $\langle S^z_i \rangle \simeq \langle S^z \rangle $, the Tyablikov equations of motion in Eq.~\ref{system} take a particularly simple form. To see this, it is convenient to introduce bosonic creation and annihilation operators, $a^+_i$ and $a^-_i=(a^+_i)^{\dagger}$ respectively, for each lattice site $i$,  which obey the standard commutation algebra
\begin{eqnarray}
[a^-_i,a^{+}_j]=\delta_{ij}, \:\:[a^{+}_i,a^{+}_j]=[a^{-}_i,a^{-}_j]=0.\label{commutators}
\end{eqnarray}
Their Heisenberg equations of motion are given by
\begin{eqnarray}
\frac{d}{dt}a_i^{\pm}&=&i[H_{\rm eff},a_i^{\pm}],\nonumber\\
H_{\rm eff}&=&-Jp\langle S^z\rangle\sum_ia^{+}_ia^-_i\nonumber\\
&&-\frac{1}{4}\langle S^z\rangle\sum_{i\neq j} J_{ij} (a^-_i-a^{+}_i)(a^-_j-a^{+}_j) \label{effective Ham}
\end{eqnarray}
where $H_{\rm eff}$ is an effective Hamiltonian that governs the time-evolution of the bosonic operators, and $J=\frac{1}{N}\sum_{i\neq j}J_{ij}\sim 9.0 J_0$ with the sum taken over all $N$ sites of the  (undiluted) square lattice. By comparing the above equation of motion with the system of equations in Eq.~\ref{system} in the coarse-graining approximation $\langle S^z_i \rangle \simeq \langle S^z \rangle $, we find that we can make the identification:
\begin{eqnarray}
\langle S^{\alpha}_i S^{\beta}_j\rangle=\langle a^{-\alpha}_ia^{-\beta}_j\rangle,
\end{eqnarray}
for all $\alpha=\pm 1$, $\beta=\pm 1$, $i,j$ and time, provided we use the same initial condition.
It now becomes clear that we can simplify the system of equations in Eq.~\ref{system} by diagonalizing the effective Hamiltonian (\ref{effective Ham}).

We make the following Bogoliubov transformation for the bosonic operators (details given in Appendix~\ref{app:bog}) which preserves the commutation relations  in Eq.~\ref{commutators}
\begin{eqnarray}
\tilde{a}^-_i &\equiv&  \sum_{j=1}^N \left [ \left[\mathbf{W}^{+ }_{i}\right]_j^{\ast} a^-_j - \left[\mathbf{W}^{-}_{i}\right]_ja^{+}_j \right ],\nonumber\\
\tilde{a}^+_i &\equiv& \left(\tilde{a}^-_i\right)^{\dagger}.
\end{eqnarray}
where $\mathbf{W^{\pm}}_i$ are $2N$-dimensional eigenvectors of the Bogoliubov matrix, defined in Appendix~\ref{app:bog}, with eigenvalues $\pm \epsilon_i$ ($i$ ranges from 1 to $N$). $\left[\mathbf{W}^{\pm}_{i}\right]_j$ is the $j$th component of the vector $\mathbf{W}^{\pm}_{i}$. The Hamiltonian becomes diagonal in the new bosonic operators $\tilde{a}^{\pm}_i$. Up to a constant
\begin{eqnarray}
H_{\rm eff}&=&2\epsilon_i \tilde{a}^{+}_i\tilde{a}^-_i,\nonumber\\
\frac{d}{dt}\tilde{a}_i^{\pm}&=&i[H_{\rm eff},\tilde{a}_i^{\pm}]=\pm 2i\epsilon_i\tilde{a}^{\pm}.\label{effective}
\end{eqnarray}
In the coarse-graining approximation, the total energy is
\begin{eqnarray}
E=NJp/4= N J p \langle S^z \rangle ^2 + \sum_{i\neq j} J_{ij} S^y_i S^y_j. \label{eq::energy}
\end{eqnarray}
Since energy is conserved by the dynamics, we can set its derivative with respect to time to zero to obtain
\begin{widetext}
\begin{eqnarray}
 \langle \dot{S}^z \rangle & = & \frac{-1}{\langle S^z \rangle 2 N J p}\frac{d}{dt}\sum_{i \neq j} J_{ij} \left\langle S^y_i S^y_j\right\rangle  = \frac{1}{\langle S^z \rangle 8 N J p}\frac{d}{dt}\sum_{i \neq j} J_{ij} \left\langle(a^-_i-a^{+}_i)(a^-_j-a^{+}_j) \right\rangle= \frac{-1}{\langle S^z \rangle  N }\frac{d}{dt}\sum_{i } \left\langle a^+_ia^-_i\right\rangle\label{eq:Sxtime}\\
  & = & \frac{-1}{N\langle  S^z\rangle}\frac{d}{d t}\sum_{i=1}^{N}\sum_{m\neq 0, n\neq 0}^{N,N} \left \langle\left( \left[\mathbf{W}^{-}_{m}\right]^{\ast}_i \tilde{a}^-_m + \left[\mathbf{W}^{+ }_{m}\right]^{\ast}_i \tilde{a}^+_m \right )\left(\left[\mathbf{W}^{+}_{n}\right]_i \tilde{a}^-_n + \left[\mathbf{W}^{-}_{n}\right]_i \tilde{a}^+_n\right ) \right \rangle \\ 
&\equiv & -\frac{1}{N \langle S^z \rangle}\sum_{m,n,\alpha ,\beta }  c^{\alpha \beta}_{mn} \dot{C}^{\alpha \beta}_{mn}(t)\, ,\end{eqnarray} 
\end{widetext}
where the Goldstone mode is not included in the summation and $C^{\alpha \beta}_{mn}  (t)\equiv \langle \tilde{a}^{\alpha}_m \tilde{a}^{\beta}_n\rangle$ ($\alpha, \beta = \pm 1$). From Eq.~\ref{effective} it follows that
\begin{equation}
 \dot{C}^{\alpha \beta}_{mn}(t) \simeq  \langle S^z \rangle \lambda^{\alpha \beta}_{mn} C^{\alpha \beta}_{mn}(t),\label{eq:Ctime}
\end{equation}
where $\lambda^{\alpha \beta}_{mn} = i(\alpha \epsilon_m +\beta \epsilon_n)$. 

 The conservation of energy can be re-expressed as
\begin{equation}
E/2NJp = \frac{1}{8} = \frac{1}{2}\langle S^z \rangle ^2 +  \frac{1}{N}\sum_{m,n,\alpha ,\beta }  c^{ \alpha \beta}_{mn} C^{\alpha \beta}_{mn}(t) .\label{eq:conservation}
\end{equation}
If $\langle S^z \rangle $ decays to a non-zero value in the long-time limit, by Eq.~(\ref{eq:Ctime}),  the coefficients $C^{\alpha \beta}_{mn}(t)$ oscillate harmonically at long times with frequencies $\alpha \epsilon_m + \beta \epsilon_n$. In the thermodynamic limit, where the frequencies form a continuum, we expect only the zero-frequency terms in  Eq.~(\ref{eq:conservation}), with $m=n$ and $\alpha+ \beta =0$,  to survive in the long-time limit (the average of the oscillating terms decaying to zero), such that
\begin{eqnarray}
\frac{1}{N}\sum_{m,n,\alpha ,\beta }  c^{ \alpha \beta}_{mn} C^{\alpha \beta}_{mn}(t) &\stackrel{t\rightarrow\infty}{=}& \frac{1}{N}\sum_{n}\left( c_{nn}^{+-}C^{+-}_{nn} + c_{nn}^{-+}C^{-+}_{nn}\right) 
\nonumber\\
&=&
2\sum_{n \neq 0} (\mathbf{W}_n^+\cdot \mathbf{W}^{+ \ast}_n)(\mathbf{W}_n^-\cdot \mathbf{W}^{- \ast}_n) 
\nonumber\\
&\equiv& F.
\end{eqnarray}
From energy conservation, we can now deduce that the long-time limit of the mean-field is given by
\begin{eqnarray}
\langle S^z (t \rightarrow \infty) \rangle = \sqrt{\frac{1}{4}-2F}.
\label{eq:effectivetheory}
\end{eqnarray}
From the above expression, it follows that $\langle S^z(t \rightarrow \infty) \rangle \neq 0$ is only possible if $F<\frac{1}{8}$. Otherwise $\langle S^z \rangle$ decays to zero at long times, in which case, by Eq.~(\ref{eq:Ctime}), all $C^{\alpha\beta}_{mn}(t)$ are constant.

This effective theory shows a mean-field-like transition at a critical value of $F=\frac{1}{8}$, assuming that  the original coupled equations lead to relaxation of $\langle S^z\rangle$. The quantity $F$ measures the degree of quantum fluctuations in the initial state. It must be computed for the specific geometry of the $N$ coupled spins. It is therefore a function of the filling $p$ (i.e of the disorder), so  $F(p_c)=1/8$ defines a critical filling $p_c$. Below the critical filling  $p<p_c$  ($F>1/8$) this theory suggests that the long-time average $\langle S^z\rangle$ will vanish in the thermodynamic limit. However, for $p>p_c$ ($F<1/8$) it suggests that  $\langle S^z\rangle$  should tend to a non-zero value. [This value grows as $\langle S^z\rangle \propto (p-p_c)^{\frac{1}{2}}$, since $F$ is linear in $p$ close to $p_c$.]
This persistence of non-zero polarization to long times is the signature of the emergent phase-coherent dynamics of the $N$ quantum spins, with the coherence stabilized against the disorder ($p\neq 1$) by the long-range dipolar coupling of the spins. Note that the existence of this stable phase  $F<1/8$ at $p>p_c$  is a special feature of our model of dipolar interactions in 2D. 
In lower dimensions, or if the interactions are nearest-neighbour only, the quantity $F$ is not bounded, so $F\gg 1/8$: in those situations, the theory predicts quantum fluctuations to lead to $\langle S^z\rangle \to 0$ at long times, that is decay of coherence of the initial state for all fillings $p$.

\subsection{Numerical results}  

\label{numerical}

We now turn to describe the results of  numerical calculations of the RPA-decoupled dynamics of our full theory, Eqns.~\ref{eq:theory},  {\it i.e.} without coarse-graining and with the time evolution of the diagonal elements of the correlation functions $C^{\alpha \beta}_{mn}(t)$ computed exactly.  Our numerical results involve studies of finite square lattices with $N_{\rm site}$ sites, and with periodic boundary conditions. For site filling $p$, there are then typically $N= p N_{\rm site}$ spins, which we take to be in the range $N=100-1000$. In view of the finite system sizes, when plotting mean values such as $\langle S^z\rangle$, we further take the average over disorder realizations.

\begin{figure}
\centering
\includegraphics[width=0.95\columnwidth]{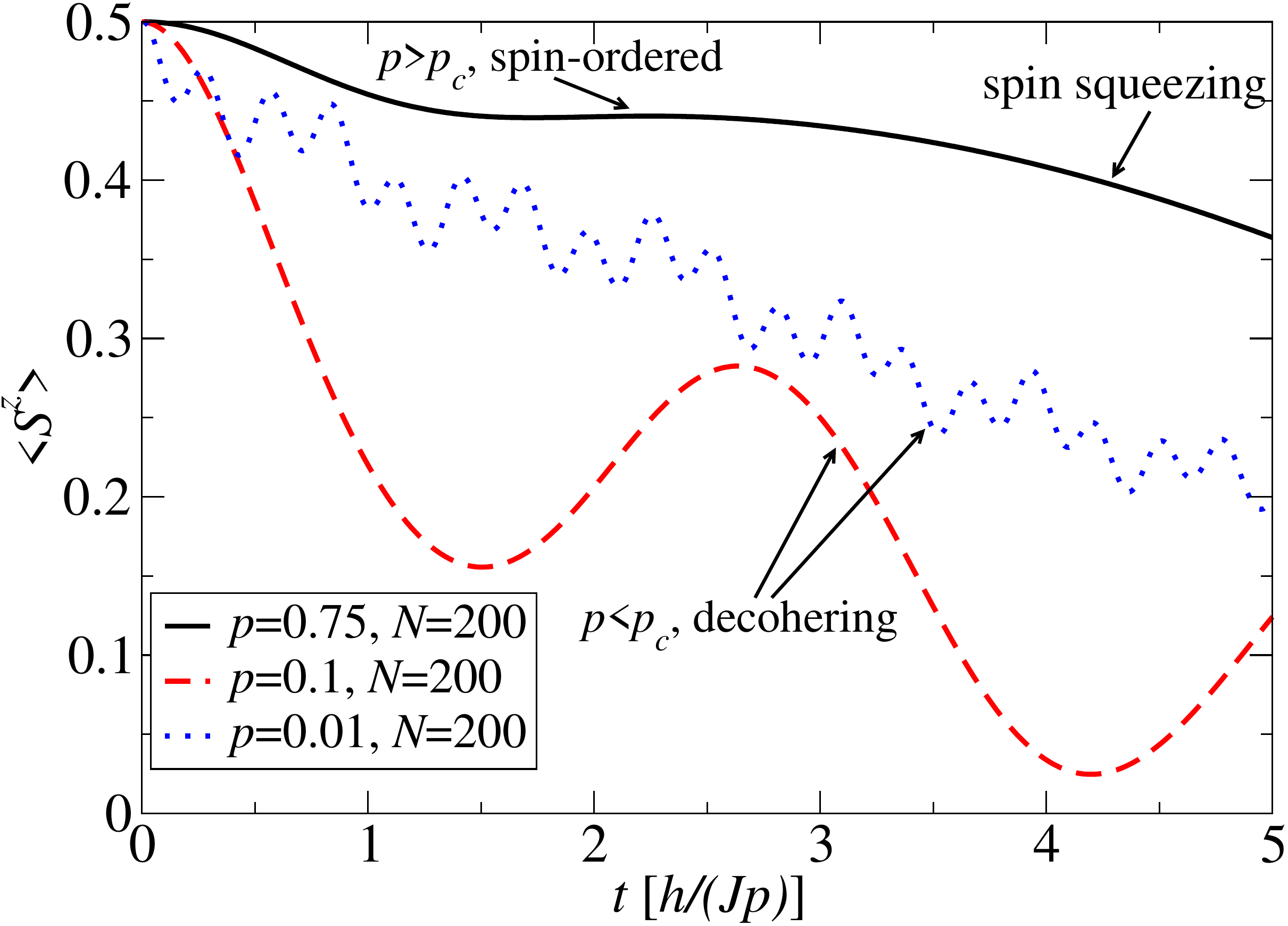}
\caption{The dynamical evolution of $\langle S^z\rangle$ for different filling fractions, $p$. For $p<p_c$, $\langle S^z\rangle$ decays to zero at long times, showing the decoherence expected for the spin-disordered phase. Decoherence occurs on a timescale  $T_2 \propto p^{-3/2}$ [see Fig.~\ref{TimeConstant}].
For $p>p_c$,  $\langle S^z\rangle$ falls to a non-zero value, signalling the formation of a phase with persistent spin-order. At very long times, $t\gtrsim \tau_{\rm sq}\propto \sqrt{N}$, 
 $\langle S^z\rangle$ begins to decay also in the spin-ordered regime $p>p_c$ as a result of spin-squeezing. }
 \label{goldstone}
\end{figure}

Fig.~\ref{goldstone} shows the time-dependence of the mean polarization $\langle S^z\rangle$ for a range of different site fillings $p$, and different system sizes $N$, in terms of time as measured in units of the rescaled mean-field coupling $Jp$.
These numerical results reproduce the qualitative features of the coarse-grained theory described above. 
 (In fact, we also find quantitatively similar results.)   In particular, the results show that there are indeed two regimes of long-time dynamics  separated by a critical filling fraction, which we find to be close to $p_c = 0.15$.

\begin{figure}
\centering
\includegraphics[width=0.95\columnwidth]{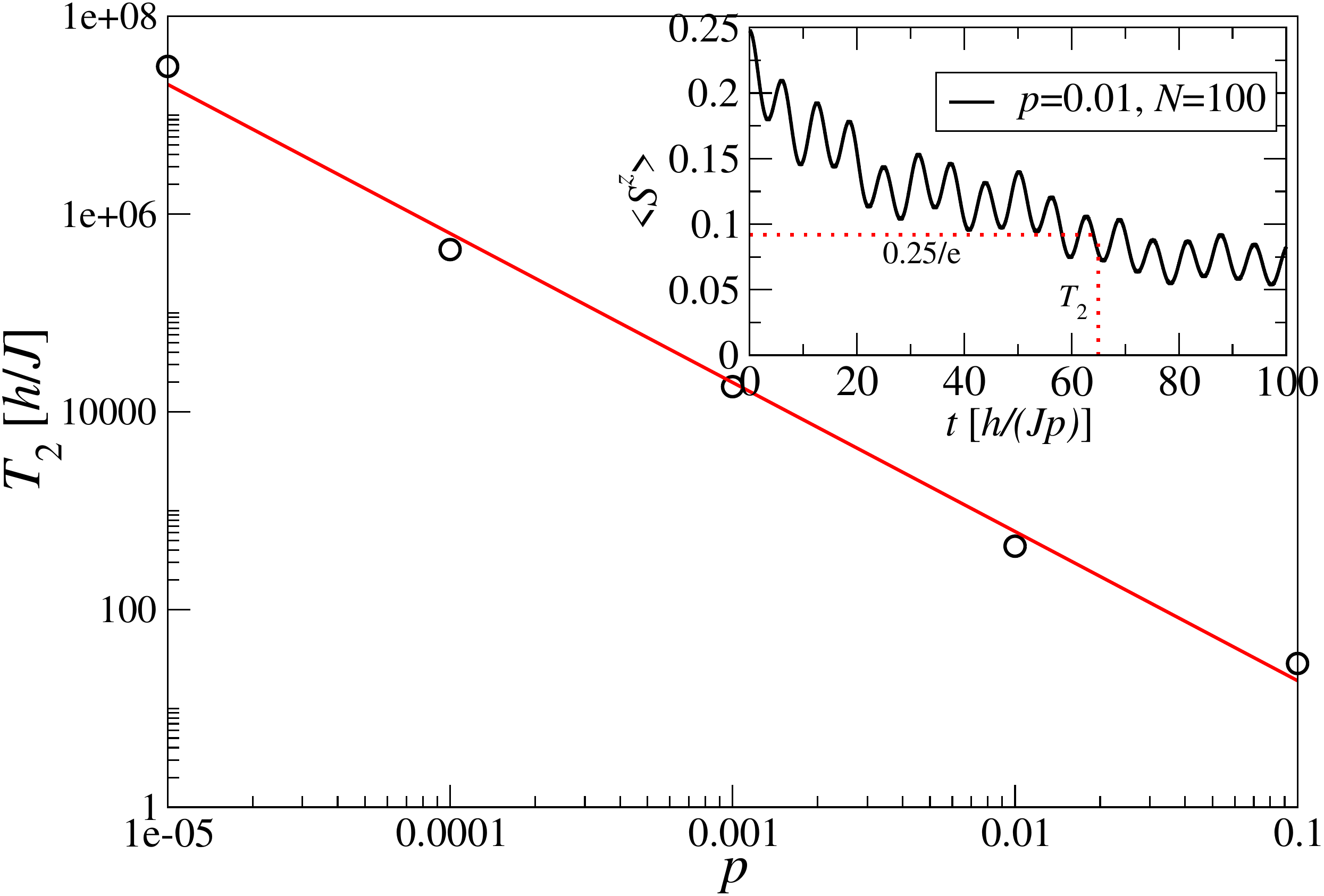}
\caption{The decoherence time $T_2$ is finite in the independent clusters regime when $p<p_c \sim 0.15$ and scales as $p^{-3/2}$ at small $p$ (solid line). The inset indicates how $T_2$ is extracted from one data set.}
\label{TimeConstant}
\end{figure}

 For $p<p_c$ the dynamics is that of independent clusters of several spins, and  $\langle S^z\rangle$ decays to zero at long times with lightly damped oscillations.  For low enough $p$, the continuum limit can be taken with the only characteristic lengthscale being the mean spacing between molecules, $p^{-1/2}a$. This lengthscale then sets both the energy and the time scale, the latter being given by $ h/(Jp^{3/2})$.  Fig.~\ref{TimeConstant} shows the coherence time as a function of $p$ in this regime $p<p_c$, and a comparison to the expected $p^{-3/2}$ scaling. 
This independent-cluster behaviour was also found in Ref.~\cite{ColoradoTheory}, which studied similar models but in regimes where disorder is always strong, and for which there is no spin-locked phase of the type we now discuss at $p>p_c$.

	Above the critical filling fraction $p> p_c$ we find that the dynamics of  the system is consistent with  relaxation, on a time scale of the order $h/(Jp)$, to a spin-ordered phase that retains long-range correlations between the spins. 
	
	For a thermodynamically large system, $N\to \infty$, this state can be characterized by a non-zero mean polarization $\langle S^z\rangle$ that persists to arbitrarily long times. This phase-locked state leads to a divergent $T_2$ decoherence time, and a persistent Ramsey interference signal. 

	In systems of finite size -- including our numerical simulations (on $N=100-1000$ spins) -- further care is required in relating the spin-order state at $p>p_c$  to the dynamics of the mean polarization $\langle S^z \rangle$. The reason is that, in any finite system, a new timescale appears, $\tau_{\rm sq} \propto \sqrt{N}$, beyond which  $\langle S^z\rangle$ tends to zero due to  the development of strong spin-squeezing.  This spin-squeezing is a result of the fact that the exact energy eigenstates are also eigenstates of total $S^x$. The differences in the energy eigenvalues of these eigenstates are proportional to $(S^x)^2 /N$ in the harmonic approximation around $S^x=0$. 
	The initial state~(\ref{eq:prepare}) is a coherent state, formed from a superposition of these eigenstates centred on $S^x=0$ with a spread that scales as $\delta S^x \propto 1/\sqrt{N}$.  The Hamiltonian $(S^x)^2 /N$ leads to one-axis squeezing of this coherent state on a timescale $\tau_{\rm sq}\propto \sqrt{N}$, along with revivals of the initial (unsqueezed) coherent state at times $\propto N$\cite{kitagawaueda}.  Indeed, our numerical results, Fig.~\ref{goldstone}, show that in the regime $p>p_c$, following an initial drop of the mean polarization $\langle S^z \rangle$ to a large non-zero value after a time  of order $(h/Jp)$, this non-zero value starts to decay away to zero at longer times, 
	 $t>\tau_{\rm sq}$. This longer timescale increases with system size in a manner consistent with $\tau_{\rm sq} \propto \sqrt{N}$ (data not shown).

\begin{figure}
\centering
\includegraphics[width=0.95\columnwidth]{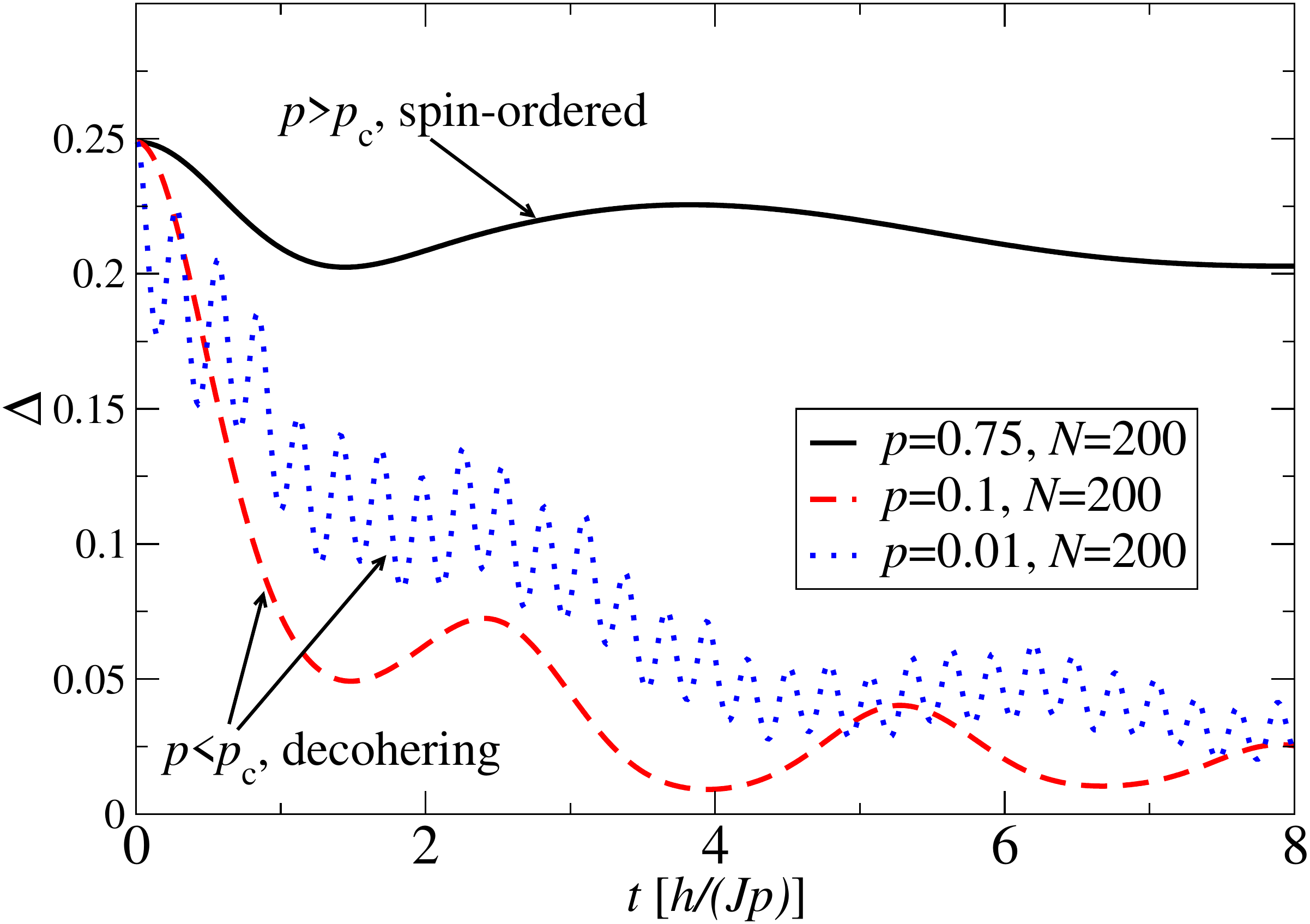}
\caption{The dynamical evolution of $\Delta$ for the same parameters as Fig.~\ref{goldstone}. For the spin-ordered phase, $p>p_c$, the order parameter remains non-zero at late times.}
 \label{fig:rho-evolution}
\end{figure}

\begin{figure}
\centering
\includegraphics[width=0.95\columnwidth]{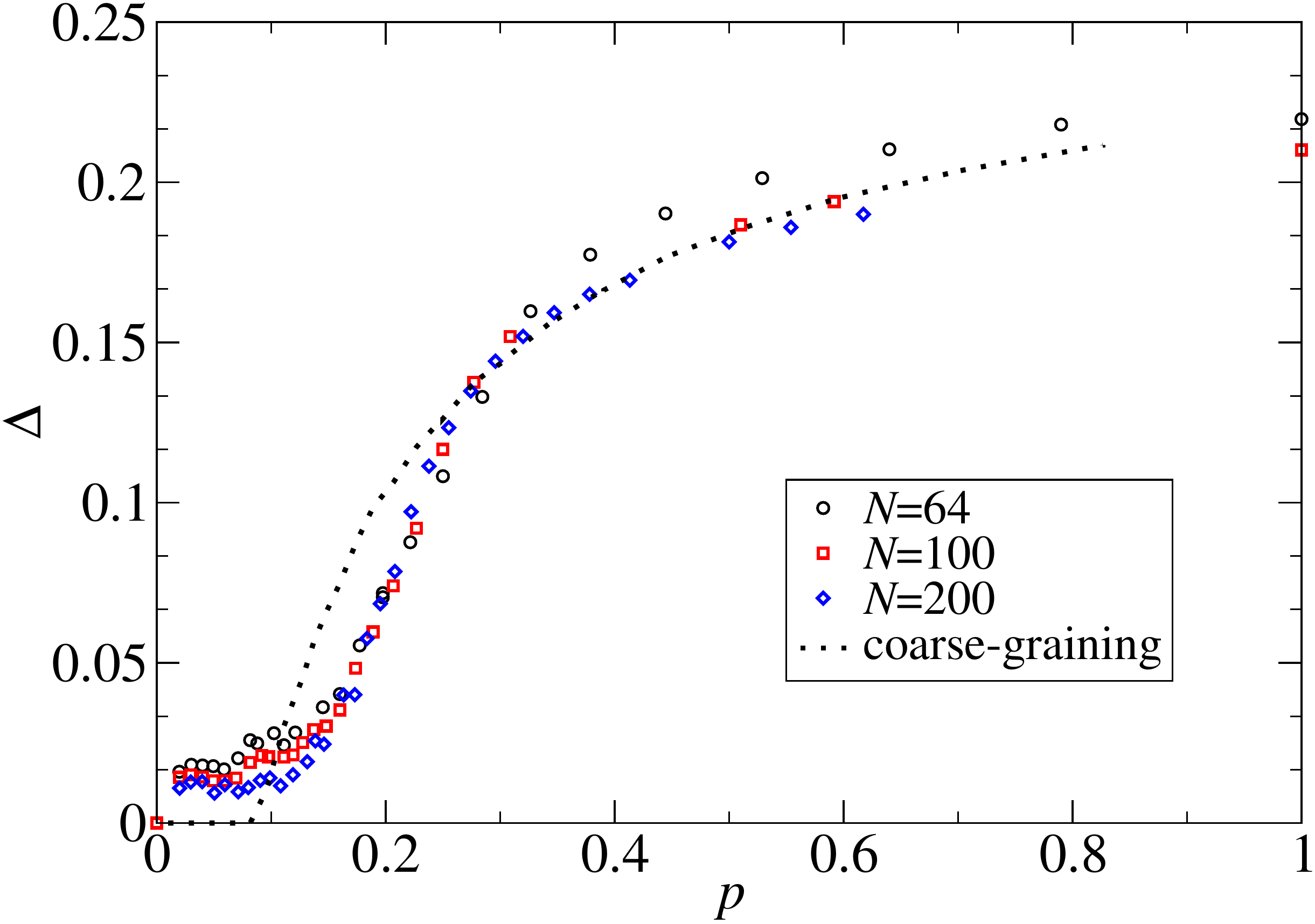}
\caption{Long-time average of the order parameter $\Delta $ as a function of the filling fraction $p$, computed from the numerical integration of the RPA equations of motion, Eqns.~\ref{eq:theory}. (The results are averaged over 100-1000 disorder realizations.) For increasing numbers of spins, $N$, the dependence on $p$ becomes increasingly sharp around $p_c\simeq 0.15$. 
  For $p<p_c$ the order parameter $\Delta$ tends to zero at late times, indicating a spin-disordered phase that decoheres. For $p>p_c$, the order parameter $\Delta$ remains non-zero for
arbitrarily long times, indicating a 
spin-ordered phase with persistent spin-coherence.
The dotted line shows the results from the analytic theory based on coarse-graining, Eq.~\ref{eq:effectivetheory}.} \label{TyablikDyn}
\end{figure}

The emergence of a collective spin-ordered phase for $p>p_c$ is most clearly observed by studying not $\langle S^z \rangle$ but the mean-square spin $\Delta$, (Eq.~\ref{eq:rho0}). This quantity reflects the existence of an infinite-range collectively ordered state of all of the spins, even in the presence of strong spin-squeezing that causes $\langle S^z\rangle$ to vanish.
In Fig.~\ref{fig:rho-evolution} we show the time evolution of the order parameter $\Delta$ for the same parameters as in Fig.~\ref{goldstone}.  In the spin-ordered phase, $p>p_c$, the order parameter $\Delta$ remains non-zero at late times.
The long-time limit of the order parameter $\Delta$, shown in Fig.~\ref{TyablikDyn}, exhibits a  transition from a 
spin-disordered phase ($\Delta\simeq 0$) at $p<p_c$, and the collective dipole-locked phase at $p>p_c$ ($\Delta \neq 0$), which becomes increasingly sharp in the thermodynamic limit, $N\to \infty$.

\section{Discussion}

Figure ~\ref{TyablikDyn} contains the main results of this
work. It shows that, for fillings $p$ above a critical value
$p_c\simeq 0.15$, the long-time dynamics  is to a
collective spin-ordered state in which all two-level systems remain phase-locked.

{\it Thermalization Ansatz:} A natural way to try to understand the long-time dynamics is to assert that the system evolves to a thermal state, that is a
state of maximum entropy constrained only by the values of any
constants of the motion. For our model, the two conserved quantities
are the energy and the mean number of excited states, which are
$E=NJp/4$ and $\rho=(1/2) p$ for the initial state we consider
(prepared by a $\theta = \pi/2$ pulse). To determine the predictions
of this thermalization ansatz for the long time polarization
$\langle S^z \rangle$ of an infinite system, we have computed the
correlation functions of the system at non-zero temperature within RPA by
considering the equation of motion of the advanced Green's function
and using Tyablikov decoupling to form a closed system of
equations. (Since the initial state is prepared with an energy
density above the mean energy density of the band, the temperature is negative\cite{kwasigrochcooper}.)
The methodology is presented in Appendix~\ref{app:thermo}.
Fig.~\ref{RPAThermo} summarizes the results of this thermodynamic analysis. This shows the equilibrium phase diagram as a function of
the mean dipolar energy per particle $E/N$ and the site occupation $p$, as
 well as the energy $E=NJp/4$ of the initial state~(\ref{eq:prepare}). These results show that, within the assumption that the system thermalizes at long times, 
there would be a transition from a disordered phase,  $\langle S^z\rangle =0$, to
a collectively ordered phase, $\langle S^z\rangle \neq 0$, at a critical filling $p_c^{\rm th} \simeq 0.15$. Hence,  we find that the critical  filling for the transition observed in the long-time dynamics, $p_c$, closely matches the value that we obtain from the thermalization ansatz, $p_c^{\rm th}$. This lends strong support to the notion that the long time dynamics can be understood in terms of the thermal equilibration of the system, and the existence of this thermodynamic phase transition. Note again that the existence of this stable ordered phase  $\Delta\neq 0$ at nonzero temperatures in this 2D setting relies on the fact that the dipole coupling is sufficiently long-ranged to suppress the phase fluctuations that preclude a magnetically ordered phase at any non-zero temperature for short-range couplings.
\begin{figure}
\centering
\includegraphics[width=0.95\columnwidth]{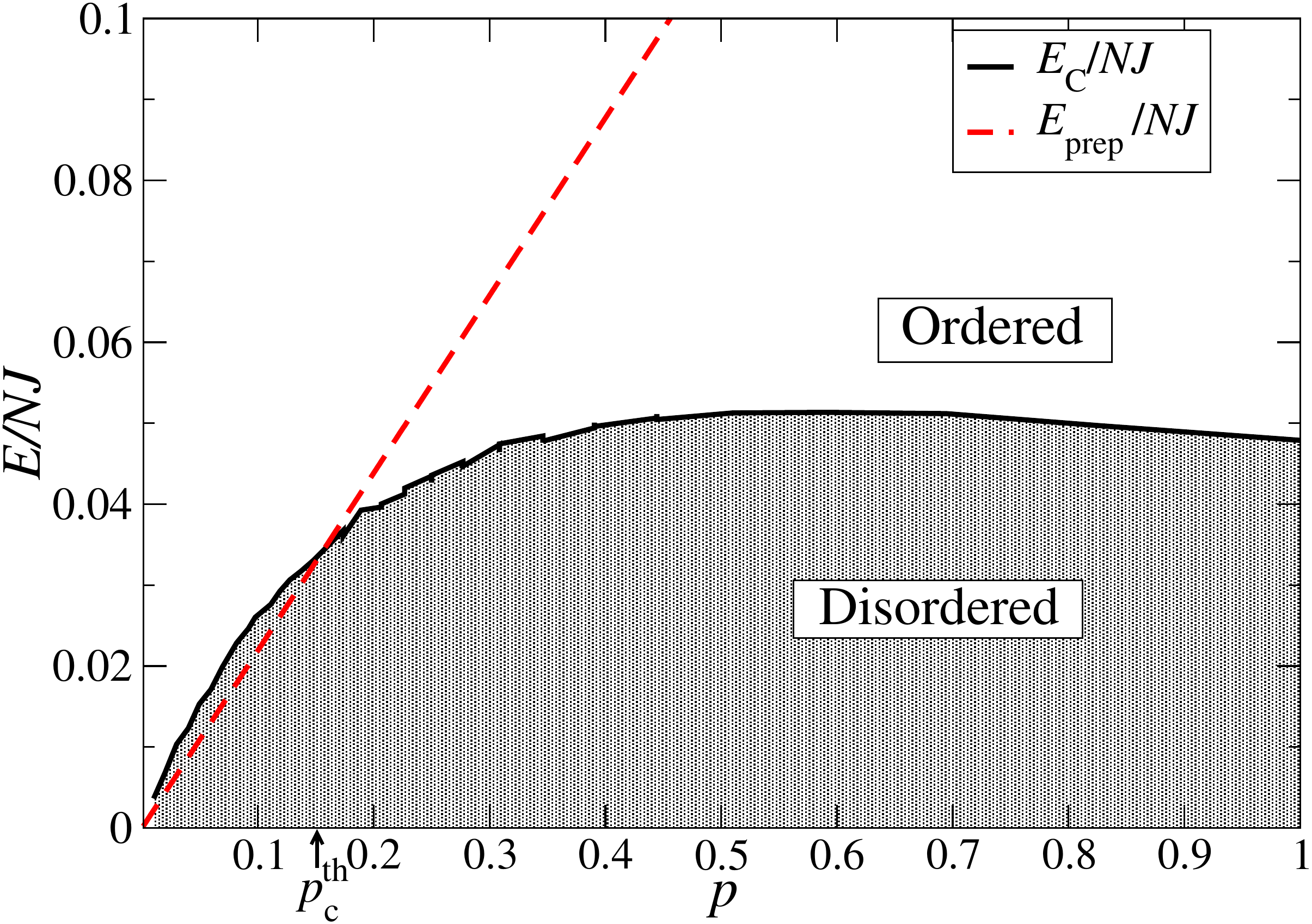}
\caption{Phase diagram  shown as a function of site dilution $p$ and mean energy per particle $E/N$. The solid line shows the critical energy, $E_c$, that determines the phase boundary between disordered and ordered phases, as computed within the equilibrium RPA theory. The dashed line is the mean energy of the initial state (\ref{eq:prepare}). Assuming thermalization of this initial state at long times, one would expect a transition from spin-disordered (i.e. decoherence) at $p < p_c$  to spin-ordered (i.e. persistent coherence) 
at $p> p_c$ with $p_c\simeq 0.15$. This expectation, based on thermodynamics,   agrees very well with the long-time behaviour found from the dynamical calculations. (See Fig~\ref{TyablikDyn}.)} \label{RPAThermo}
\end{figure}

\begin{figure}
\centering
\includegraphics[width=0.95\columnwidth]{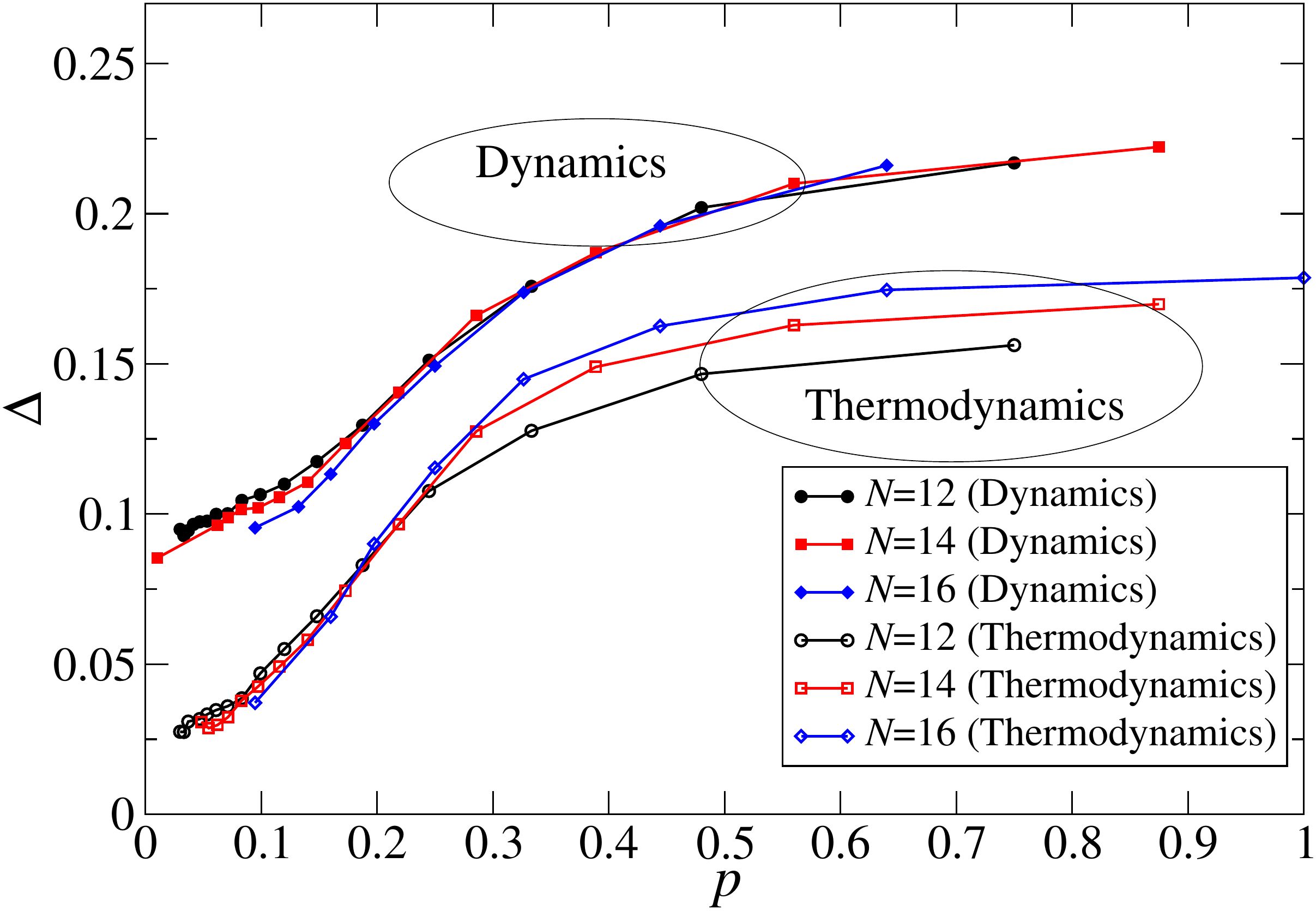}
\caption{Long-time average of $\Delta$ calculated from exact dynamics of small systems compared against the corresponding thermodynamic prediction of the microcanonical ensemble. There is a cross-over in system-size scaling for both at approximately $p\sim 0.2$ which agrees well with the RPA results} \label{EDDynamics}
\end{figure}

{\it Exact Diagonalization Studies:} To explore the properties of this strongly interacting many body quantum system beyond the above approximate schemes, we have performed  exact diagonalization studies on finite systems with periodic boundary conditions. Our results are restricted to small systems, with a number of spins in the range $N=8-16$ spins.  To suppress fluctuations arising from the specific disorder realization, we average the results over 1000 disorder realizations for the smaller systems and 100 disorder realizations for the larger systems. 
 We have used the exact diagonalization studies to test the results obtained using the RPA decoupling scheme in both the dynamical and thermodynamical properties. To do so, we calculate $\Delta$ computed in two different ways. For the dynamical case, we compute the expectation value using the state obtained by following the exact time-evolution of the initial state to the long-time limit.  For the thermodynamic case, we use the expectation value in a microcanonical ensemble, in which we average over the five eigenstates that are closest in energy to the mean energy of the prepared state. The results of these two methods are summarized in Fig.~\ref{EDDynamics}.  They show qualitatively the same behaviour of $\Delta$ calculated either from 
the long-time dynamics or from the  thermodynamic average: in both cases, $\Delta$ shows a marked rise from a small to large values as $p$ increases through a value $\simeq 0.15$. We associate this crossover in these small finite systems with the phase transition found in the RPA approach for $N\to \infty$ at a value of $p_c\simeq 0.15$. That both the dynamical and thermodynamical results agree in capturing the behaviour in the vicinity of the transition $p\simeq 0.15$ is consistent with the results we found in the RPA decoupling. This implies that the eigenstate thermalization hypothesis (ETH)\cite{Huse2013} is accurate for states in this regime of energy.

{\it Validity of Mean-Field Methods:} It should come as no surprise that our mean-field type theories do rather well. The disordering transition can be thought of as a thermodynamic phase transition taking place at an effective non-zero temperature. The thermal excitations that dominate and drive this transition consist of long-wavelength spin-waves in which the discrete nature of the participating microscopic spins becomes coarse-grained. The system flows, in an RG sense, to an effective large-spin description at large lengthscales. This forms the basis of the semi-classical spin-wave analysis. Furthermore, the effective Ginzburg-Landau energy describing the phase transition can be written as
\begin{equation}
F= \int d^d \mathbf{r} \left\{ t |\Psi(\mathbf{r})| ^2 + \frac{K}{2}\left|\nabla^{\frac{1}{2}} \Psi(\mathbf{r})\right|^2 +u  |\Psi(\mathbf{r})|  ^4\right\},
\label{eq:gl}
\end{equation}
 where $\Psi(\mathbf{r})$ is the coarse-grained magnetization and  $\nabla^{\frac{1}{2}} \Psi(\mathbf{r})$ is a consequence of the anomalous relativistic dispersion. One could view this effective Ginzburg-Landau functional as originating from the microscopic self-consistency equations (see Eqn~\ref{eq:recursion}). As a result of the anomalous, relativistic dispersion, the upper critical dimension of the theory is 2, giving only logarithmic corrections to mean-field exponents in $D=2$. (See Appendix~\ref{app:upper}.)

\section{Summary}

We have studied the decoherence dynamics of dipole-coupled two-level systems in 2D,  subjected to Ramsey-type experiments. The combination of $r^{-3}$ power-law interactions and the dimensionality, $D=2$, allows the appearance both of a spin-ordered phase at non-zero temperature, and a spin-disordered phase at sufficiently strong disorder. We studied the transition between these two phases for the situation in which the disorder arises from site dilution, with a  fraction $p$ of the lattice sites hosting  a two-level system. This  is the dominant form of disorder in the experimental realization provided by polar molecules trapped in a deep optical lattice\cite{ColoradoExp}.
We developed a self-consistent mean-field theory that is capable of describing the effects of this  form of disorder. For small site occupations, the long-time dynamics shows decoherence with $T_2\propto p^{-3/2}$ set by the (inverse of the) dipole interaction energy at the mean interparticle spacing $\sim p^{1/2} a$. This is similar to behaviour in existing experimental studies of decoherence dynamics of rotational levels of polar molecules\cite{ColoradoExp,ColoradoTheory}, with an increase in the filling, $p$, causing a decrease in  $T_2$ (in that case, of a 3D setting,  the mean interparticle spacing is $\sim p^{1/3} a$ so $T_2\propto p^{-1}$).   Our results show that, in stark contrast to this, increasing the filling $p$ can in fact {\it increase} the degree of coherence. Indeed, for $p$ larger than a critical value $p_c\simeq 0.15$ we predict that a 2D system  will show  a transition to a spin-ordered phase for which, for a large system $N\to \infty$, phase coherence is retained for arbitrarily long times, i.e.  persistent Ramsay oscillations with divergent $T_2$ decoherence time. For a finite system composed of $N$ two-level systems, this collective spin-ordered phase begins to exhibit
spin-squeezing for times longer than a timescale $\tau_{\rm sq}\propto \sqrt{N}$. Evidence of the spin-ordered phase can continue to be found in an order parameter $\Delta$, determined by the variances of the total spin. The critical occupation that we predict, $p_c\simeq 0.15$, is already achievable in existing experiments on polar molecules in 3D lattices\cite{lowentropymolecules,innsbruck}.
Our results present the exciting possibility that Ramsey-type experiments on 2D planes in these systems could be used to explore this novel phase transition in decoherence dynamics, 
from a disorder-dominated regime at $p<p_c$ where dipole interactions cause decoherence, to  a spin-ordered phase at $p>p_c$ in which the dipole interactions stabilize long-range coherence even in the presence of disorder.

We conclude by noting that, although our study has been motivated by
the dipolar interactions of the rotational levels of polar molecules,
similar physics can arise in other settings with long-range coupled
two-level systems. For a model with $1/r^\alpha$ interactions in
$D$ dimensions, the conditions for the existence of a well-defined
thermodynamic limit and stable localized states is $\alpha > D$, while
the condition for an ordered (BEC) phase at non-zero temperature is
$\alpha < 2D$.  Thus, we anticipate similar physics to appear for
$\alpha/2 < D < \alpha$. Such settings may be realizable in certain trapped-ion
quantum simulators\cite{porrascirac,Britton2012}.

\appendix

\section{Bogoliubov diagonalization}

\label{app:bog}

 We largely follow the work of Ref.~\cite{Rossignoli} here, which should be consulted for derivations of any statements that are quoted without a detailed proof.
The effective Hamiltonian introduced in Eq.~\ref{effective} can be rewritten as follows
\begin{eqnarray}
&&H_{\rm eff}/\langle S^z\rangle =\nonumber\\
&&=-2Jp\sum_ia^{+}_ia^-_i-\frac{1}{4}\sum_{i\neq j} J_{ij} (a^-_i-a^{+}_i)(a^-_j-a^{+}_j) \nonumber\\
&&\equiv A_{ij}(a^{+}_ia^-_j+a^-_ia^{+}_j)+B_{ij}(a^-_ia^-_j+a^{+}_ia^{+}_j)  \label{eq::Hamiltonian}
\end{eqnarray}
where we have neglected an overall constant. 
We define a Bogoliubov transformation to a new set of bosonic creation/annihilation operators $\tilde{a}^{\pm}_i$
\begin{eqnarray}
\tilde{a}^-_i &\equiv&  \sum_{j=N+1}^{2N}  \left[\mathbf{W}^{- }_{i}\right]_j a^-_j - \sum_{j=1}^{N} \left[\mathbf{W}^{-}_{i}\right]_ja^{+}_j ,\nonumber\\
\tilde{a}^+_i &\equiv& -\sum_{j=N+1}^{2N}  \left[\mathbf{W}^{+ }_{i}\right]_j a^-_j + \sum_{j=1}^{N} \left[\mathbf{W}^{+}_{i}\right]_ja^{+}_j ,
\end{eqnarray}
where $\mathbf{W}_i^{\pm}$ are a family of $2N$-dimensional vectors indexed by $i$ ranging from 1 to $N$.
These vectors satisfy the following properties
\begin{eqnarray}
(\mathbf{W}^{\pm}_i)^{\dagger} \begin{pmatrix} 1 && 0 \\ 0 && -1 \end{pmatrix}\mathbf{W}^{\pm}_j=\pm \delta_{ij},\label{properties1}\\
\mathbf{W}^-_n= \begin{pmatrix} 0 && 1 \\ 1 && 0 \end{pmatrix} (\mathbf{W}^+_n)^{\ast},\label{properties2}
\end{eqnarray}
where the first property is a consequence of bosonic commutation relations of the new operators $[\tilde{a}^-_i,\tilde{a}^+_j]=\delta_{ij}$, and the second one is a consequence of their hermitian conjugation $\tilde{a}^+_i=\left(\tilde{a}^-_i\right)^{\dagger}$.
Straightforward matrix manipulation shows that the Hamiltonian in Eq.~\ref{eq::Hamiltonian} is diagonal in the new bosonic operators
\begin{eqnarray}
H_{\rm eff}/\langle S^z\rangle = \sum_{n\neq 0} 2\epsilon_n (\widetilde{a}^{\dagger}_n \widetilde{a}_n+\frac{1}{2}) \label{DiagHam}
\end{eqnarray}
provided that the vectors $\mathbf{W}_i^{\pm}$ are eigenvectors of the matrix
\begin{eqnarray}
\mathcal{H}= \begin{pmatrix} A_{ij} && B_{ij} \\ -B_{ij} && -A_{ij} \end{pmatrix}.
\end{eqnarray}
 Their eigenvalues are $\pm\epsilon_n$ respectively. It turns out that such diagonalization is always possible because the eigenvectors of $\mathcal{H}$ automatically satisfy the properties in Eq.~\ref{properties1} and Eq.~\ref{properties2} (see Ref.~\cite{Rossignoli} for further details). (The second property is satisfied if the eigenvalues of $\mathcal{H}$ are real, which is the case here because we are expanding about the classical ground state). Note that, since the Hamiltonian (\ref{Ham}) has an exact symmetry under global rotations about the $z$-axis, one of the modes $n=0$ is a Goldstone mode, with $\epsilon_{n=0}=0$. This zero-energy mode is excluded from the sum over simple-harmonic oscillators in Eq.~\ref{DiagHam}. 
 
\begin{widetext}

\section{RPA Thermodynamic Analysis}

\label{app:thermo}

Following the work of Ref.~\cite{Tyablikov}, we begin with the advanced Green's function 
\begin{eqnarray}
G_{ij}^{\alpha \beta} (t)\equiv i \Theta (-t) \langle [S^{\alpha}_i (t), S^{\beta}_j(0)]\rangle_T.
\end{eqnarray}
By differentiating both sides of the above equation with respect to time and subsequently Tyablikov-decoupling higher-order Green's functions, e.g. $i \Theta (-t)\langle[S^z_k (t) S^+_i(t),S^-_j(0)]\rangle_T \simeq \langle S^z_k\rangle G^{+ -}_{ij}(t)$, the resulting RPA equations of motion can be written in matrix form
\begin{equation}
\begin{pmatrix} E\mathbf{1} + \mathbf{A}^{\prime} & \mathbf{B}^{\prime}  \\ -\mathbf{B}^{\prime} & E\mathbf{1} - \mathbf{A}^{\prime} \end{pmatrix} \begin{pmatrix}\mathbf{G}^{+ - }(E)  &\mathbf{G}^{+ +}(E) \\ \mathbf{G}^{- -}(E) & \mathbf{G}^{- +}(E)\end{pmatrix} \equiv (E \mathcal{I} + \mathcal{H}^{\prime}) \mathcal{G}(E)= 
\begin{pmatrix}\mathbf{S}^z & 0 \\ 0 & -\mathbf{S}^z\end{pmatrix}
\end{equation}
where the elements of the matrices $\mathbf{A}^{\prime}$, $\mathbf{B}^{\prime}$ and $\mathbf{S}^z$ are given by 
\begin{eqnarray}
 && [\mathbf{A}^{\prime}]_{ij}=4 \delta_{ij}\sum_a J_{ia} \langle S^z_a\rangle-2J_{ij}\langle S^z_i\rangle,
[\mathbf{B}^{\prime}]_{ij}=2J_{ij}\langle S^z_i\rangle,
\nonumber\\
&& [\mathbf{S}^z]_i= \langle S^z_i \rangle_T ,
\end{eqnarray}
$\mathbf{1}$ and $\mathcal{I}$ are $N$-dimensional and $2N$-dimensional identity matrices respectively, and where $G^{\alpha \beta}_{ij}(t) = \frac{1}{2 \pi}\int^{+\infty}_{-\infty} G^{\alpha \beta}_{i j}(E) \exp^{-iEt}dE$. Working in the eigenbasis of $\mathcal{H}^{\prime}$ we write down $N$ self-consistent equations for $\langle S^z_i \rangle$ using the fluctuation-dissipation theorem
\begin{eqnarray}
\langle S^z_i\rangle &=& \frac{1}{2} - \langle S^-_iS^+_i\rangle \\
&=&\frac{1}{2} - \int^{+\infty}_{-\infty} \frac{\frac{-1}{\pi}\Im \left[\mathcal{G}^{-+}_{ii}(E-i0^+)\right ]}{e^{\beta E}-1} dE\\
&=& \frac{1}{2} + \left[\left(e^{-\beta \mathcal{H}^{\prime}}-1\right)^{-1}   \right]^{- +}_{ii}\langle S^z_i\rangle.\label{eq:recursion}
\end{eqnarray}
This forms a system of $N$ coupled equations that can be solved recursively. However, we find that making the coarse-graining approximation $\langle S^z_i\rangle \simeq \langle S^z \rangle$, $\mathcal{H}^{\prime} \simeq -8\langle S^z\rangle\mathcal{H}$ is much less numerically demanding and has a negligible impact on the results. Within this approximation the critical temperature at which the magnetization vanishes can be extracted from Eq.~\ref{eq:recursion} by linearizing it in $\langle S^z\rangle$
\begin{eqnarray}
T_c & =& -\frac{4N}{ \sum_i [\mathcal{H}^{-1}]^{-+}_{ii}} 
 = -\left[\sum_{i \neq 0} \frac{4}{N \epsilon_i}(\mathbf{W}^{- \ast}_i\cdot \mathbf{W}^{-}_i+\mathbf{W}^{+ \ast}_i \cdot \mathbf{W}^{+}_i)\right]^{-1}\! .
\end {eqnarray}
There is no zero-temperature transition within RPA and the magnetization at zero temperature vanishes continuously as $p\rightarrow 0$
\begin{eqnarray}
S^z = \frac{1}{2(1+\frac{1}{N} \sum_{i \neq 0} \mathbf{W}^-_i \cdot \mathbf{W}^{- \ast}_i)}, 
\end{eqnarray}
where $\sum_{i \neq 0} \mathbf{W}^-_i \cdot \mathbf{W}^{- \ast}_i$ is the depletion due to zero-point spin-wave fluctuations. Finding the exact solution of Eq.~\ref{eq:recursion} recursively does not change this result significantly.

At the employed level of approximation the residue $\langle S^z_i S^z_j\rangle- \langle S^z_i\rangle\langle S^z_j\rangle$ is implicitly neglected. However, this is no longer valid in the absence of symmetry breaking and has to be included to get an accurate estimate of the energy density at the critical temperature. One could consider the equation of motion of higher order advanced Green's functions $i \Theta (-t) \langle [S^{\alpha}_i (t) S^{\beta}_j (t), S^{\alpha}_k(0)S^{\beta}_l(0)]\rangle$ to find the residue. However, in the absence of spontaneous symmetry breaking $\langle S^z_i S^z_j\rangle = \langle S^y_i S^y_j\rangle$ purely from symmetry considerations, and the energy of the system at the critical temperature $T_c$ is thus given by 
\begin{eqnarray}
\langle H\rangle_{T_c} & = & 2\sum_{i,j}J_{ij}\langle S^y_iS^y_j\rangle=-\frac{1}{2}\sum_{i,j}J_{ij}(S^+_i-S^-_i)(S^+_j-S^-_j)\\
& = & -\frac{T_c}{16}\sum_{i,j} J_{ij}\left( [\mathcal{H}^{-1}]_{ij}^{+-}-[\mathcal{H}^{-1}]_{ij}^{-+} + [\mathcal{H}^{-1}]_{ij}^{++} - [\mathcal{H}^{-1}]_{ij}^{--}\right)\\
& = & -\frac{T_c}{16} \sum_{i \neq 0}\frac{1}{\epsilon_i}(\mathbf{W}^{- \ast}_i \mathbf{J} \mathbf{W}^{-}_i+\mathbf{W}^{+ \ast}_i \mathbf{J} \mathbf{W}^{+}_i -\mathbf{W}^+_i \mathbf{J} \mathbf{W}^-_i  - \mathbf{W}^-_i \mathbf{J} \mathbf{W}^+_i+ {\rm c.c.})
\end{eqnarray}
where we have again linearized the fluctuation-dissipation theorem Eq.~\ref{eq:recursion} in $\langle S^z\rangle$.
\end{widetext}

\section{Upper Critical Dimension}

\label{app:upper}

To determine the upper critical dimension of the theory (\ref{eq:gl}), we express the order parameter as a sum of its mean-field value $\bar{m}$ and fluctuations $\delta\Psi(\mathbf{r})$
\begin{eqnarray}
\Psi(\mathbf{r}) = \bar{m} + \delta \Psi(\mathbf{r}).
\end{eqnarray}
For mean-field critical exponents to hold, the $t$-dependent mean-field contribution to the energy density must dominate over the fluctuations in the limit $t\rightarrow 0$~\cite{UpperCritical}
\begin{eqnarray}
&&\langle |\Psi(\mathbf{r})|^2\rangle \rangle = \bar{m}^2 + \langle |\delta \Psi(\mathbf{r})|^2 \rangle
\nonumber\\
&&=
  \bar{m}^2 + \sum_{\alpha=1,2}\int_{|\mathbf{q}| \lesssim\frac{1}{a}} \frac{d^D\mathbf{q}}{\left(2 \pi\right)^D} \frac{1}{\xi_{\alpha}^{-1} + K|\mathbf{q}|}
\nonumber\\
&&=
\bar{m}^2 + \sum_{\alpha=1,2}\int_{|\mathbf{q}| \lesssim\frac{1}{a}} \frac{d^D\mathbf{q}}{\left(2 \pi\right)^D} \frac{1}{K |\mathbf{q}|} 
\nonumber\\
&&
- \sum_{\alpha=1,2}\int_{|\mathbf{q}| \lesssim\frac{1}{a}} \frac{d^D\mathbf{q}}{\left(2 \pi\right)^D} \frac{\xi_{\alpha}^{-1}}{K|\mathbf{q}|\left(\xi_{\alpha}^{-1}+ K|\mathbf{q}|\right)},
\nonumber\\
\end{eqnarray}
where $\bar{m} = \sqrt{\frac{-t}{4u}}$ for $t<0$ and vanishes otherwise, and $\xi_1^{-1} = t+12u\bar{m}^2$, $\xi_2^{-1} = t + 4u\bar{m}^2$. Let us now analyse the different contributions to $\langle |\Psi(\mathbf{r})|^2\rangle$ appearing in the final line in the limit $t \rightarrow 0$, i.e. close to the critical point. The mean-field contribution scales as $\bar{m}^2 \sim t$, the first fluctuations integral is a constant, whereas the second fluctuations integral (with the UV cutoff safely removed) scales as $\xi_{\alpha}^{1-D} \sim t^{D-1}$. We thus deduce that the mean-field contribution dominates as $t\rightarrow 0 $ when $D>2$.

\acknowledgments{We are grateful for helpful discussions with Andreas Nunnenkamp and Ana Maria Rey, and for financial support from EPSRC Grant Nos EP/K030094/1, EP/P009565/1 and the Simons Foundation. }

\end{document}